\documentclass[12pt,a4paper]{article}
\usepackage{amsmath}
\usepackage[dvips]{graphicx}
\input epsf

\def\lsim{\mathrel{\rlap{\lower4pt\hbox{\hskip1pt$\sim$}}
    \raise1pt\hbox{$<$}}}                % less than or approx. symbol
\def\gsim{\mathrel{\rlap{\lower4pt\hbox{\hskip1pt$\sim$}}
    \raise1pt\hbox{$>$}}}                % greater than or approx. symbol

\newcommand{\xp}{\ensuremath{x_{I\!\!P}}}
\newcommand{\dperp}{d_{\perp}}
\newcommand{\bperp}{b_{\perp}}
\newcommand{\lperp}{l_{\perp}}

\def\lsim{\mathrel{\rlap{\lower4pt\hbox{\hskip1pt$\sim$}}
    \raise1pt\hbox{$<$}}}                % less than or approx. symbol
\def\gsim{\mathrel{\rlap{\lower4pt\hbox{\hskip1pt$\sim$}}
    \raise1pt\hbox{$>$}}}                % greater than or approx. symbol

\begin{document}

\thispagestyle{empty}

\begin{center}
{\Large {\bf The dipole picture of small-$x$ physics \\
(A summary of the Amirim meeting)}}
\vspace*{1.0cm}

M. F. McDermott$^1$

\vspace*{0.5cm}
%{\small
$^1$Department of Physics and Astronomy, University of Manchester, Manchester, 
M13 9PL, England.\\
email: mm@a13.ph.man.ac.uk
%}

%\vspace*{0.5cm}

%and 

%\vspace*{0.5cm}

%H.~Abramowicz$^2$,  J.~Bartels$^3$, L.~Frankfurt$^2$, K.~Golec-Biernat$^4$, E.~Gurvich$^2$, H.~Jung$^5$, S.~Kananov$^2$, A.~Kreisel$^2$, H.~Kowalski$^4$, E.~Levin$^2$, A.~Levy$^2$, P.~Marage$^6$,  U.~Maor$^2$,  E.~Naftali$^2$, P.~Saull$^{2,7}$, S.~Schlenstedt$^8$, G.~Shaw$^1$, M.~Strikman$^7$, K.~Tuchin$^2$, J.~Whitmore$^7$

%\vspace*{0.5cm}

%{\small
%$^2$School of Physics and Astronomy, Raymond and Beverly Sackler 
%Faculty of Exact Sciences, Tel Aviv University, Tel Aviv, Israel.\\
%$^3$ II. Institut f\"{u}r Theoretische Physik der Universit\"{a}t Hamburg, 
%Luruper Chausee 149, D-22761, Hamburg, Germany.\\
%$^4$ DESY, Notkestrasse 85, D-22603, Hamburg, Germany. \\
%$^5$ Elementary Particle Physics, Lund University, SE-221 00, Lund, Sweden.\\
%$^6$ Universit\'{e} Libre de Bruxelles, CP230, Boulevard du Triomphe, B-1050, 
%Bruxelles, Belgium\\
%$^7$Department of Physics, Penn State University, University Park, PA, USA.\\
%$^8$ DESY-Zeuthen, Platanenallee, D-15735 Zeuthen, Germany. \\
%}

\vspace*{1.0cm}

\begin{abstract}
A personal summary of the discussions which took place at 
the informal meeting in Amirim, Israel from June 1-4 2000,
concerning the dipole picture of small-$x$ physics is presented.
The broad aim of the meeting was to address the question
``Has HERA reached a new QCD regime (at small $x$) ?''.  
The new regime in question is the high-density, but weak-coupling, 
limit of perturbative QCD. 
%This summary is prepared in collaboration with experimentalists and 
%theorists that attended the meeting.
\end{abstract}

\end{center}

\newpage

\section{Introduction and background}

{}From 1-4 June 2000 approximately twenty HEP  
experimentalists and theorists met at Amirim, a small vegan 
village in Northern Israel, to discuss the question 
``Has HERA reached a new QCD regime (at small $x$) ?''. 
The meeting was organized by Tel Aviv University and was somewhat 
unusual and innovative in style in that discussions took place 
``al fresco'', at a white board in a small garden,  
and there were no formal presentations. This paper aims to 
summarize the discussions and is written with the assistance of those 
who attended the meeting\footnote{See Acknowledgments.}.

\begin{figure}[htb] 
\vspace{2mm}
\begin{center}
\leavevmode
\hbox{%
\epsfxsize=5in
\epsfysize=7.5in
\epsffile{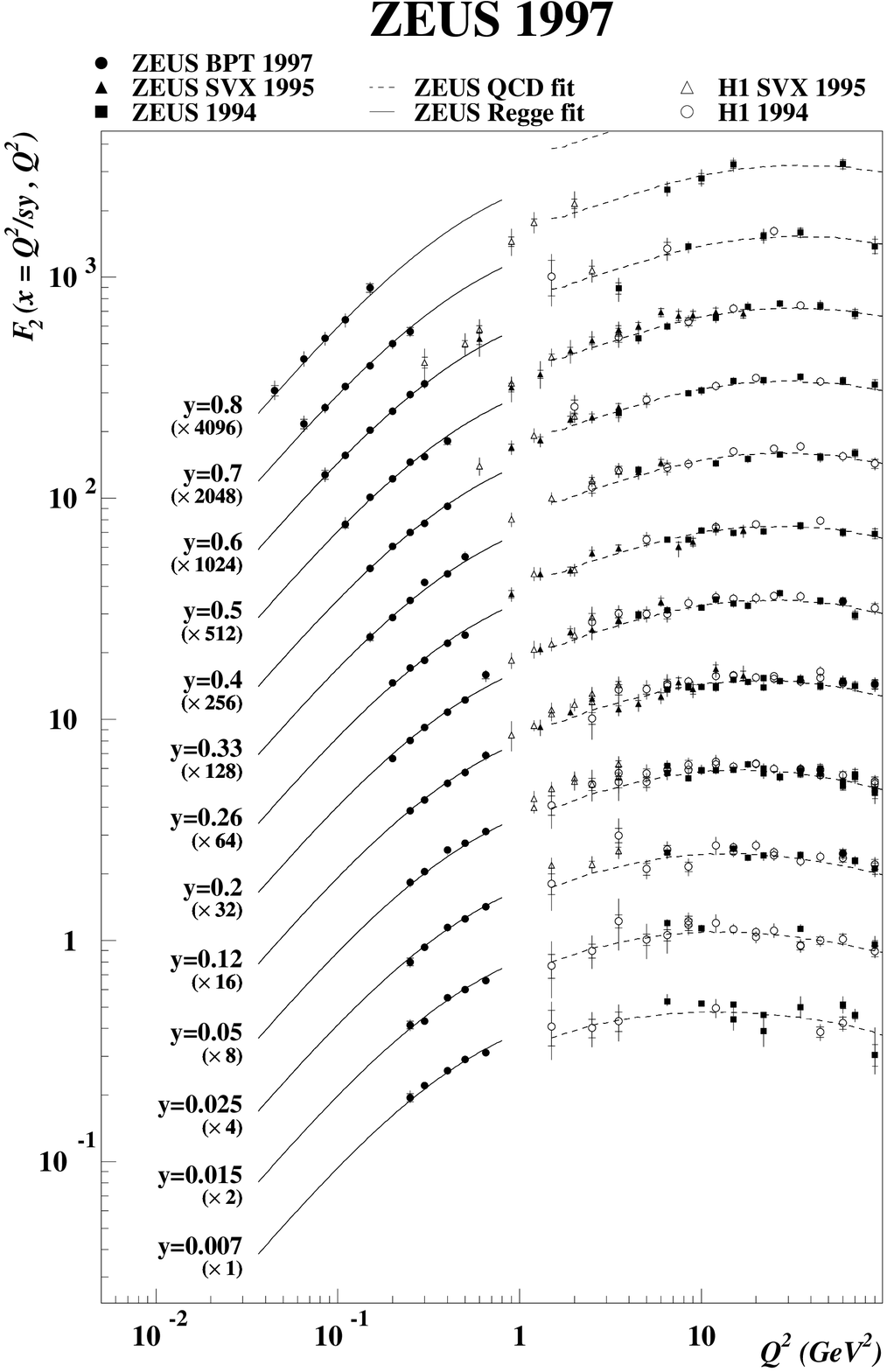}}
\end{center}
 \caption[f2]{The structure function $F_{2}(x,Q^2)$ as a function of $Q^2$. The 1997 ZEUS data \cite{zlowq} reveals the approximate $F_2 \propto Q^2 $ behaviour for the lowest values of $Q^2$.}
\label{fig:lowq} 
\end{figure}

In recent years improvements to the H1 and ZEUS detectors, have enabled 
the measurement of the Deep Inelastic Scattering (DIS) structure function, 
$F_{2} (x,Q^2)$ to be pushed to new and interesting kinematical regions in the 
photon's virtuality, $Q^2$, and scaling variable $x \approx Q^2/W^2$ 
($W$ is the centre of mass energy of the photon-proton interaction). 
In particular, the new high-precision very low $Q^2 \ll 1.0 $~GeV$^2$  data 
\cite{zlowq,hlowq,klein} allows the important transition from DIS to 
photoproduction to be studied, at low values of $x$, for the first time. 
It is known that $F_2$ must be proportional to $Q^2$ as $Q^2 \rightarrow 0 $ 
(to satisfy conservation of electromagnetic current and to achieve a 
finite photoproduction cross section) and indeed this behaviour 
appears to have been observed by ZEUS at the lowest $Q^2$ values 
(see Fig.(\ref{fig:lowq}) and \cite{zlowq}). This implies a departure 
from the logarithmic scaling violations predicted by standard perturbative 
QCD evolution analyses (leading-twist DGLAP\cite{dglap}) which are anyway 
not expected to be applicable for such low scales (the input scale $Q_0^2$ is currently 
chosen to be around $1.0 $~GeV$^2$ and all data with $Q^2 > Q_0^{2}$ are analyzed, 
implying that the higher-twist corrections to the standard picture are not important above $Q_0^2$). 

\begin{figure}[htb] 
\vspace{2mm}
\begin{center}
\leavevmode
\hbox{%
\epsfxsize=5in
\epsfysize=7in
\epsffile{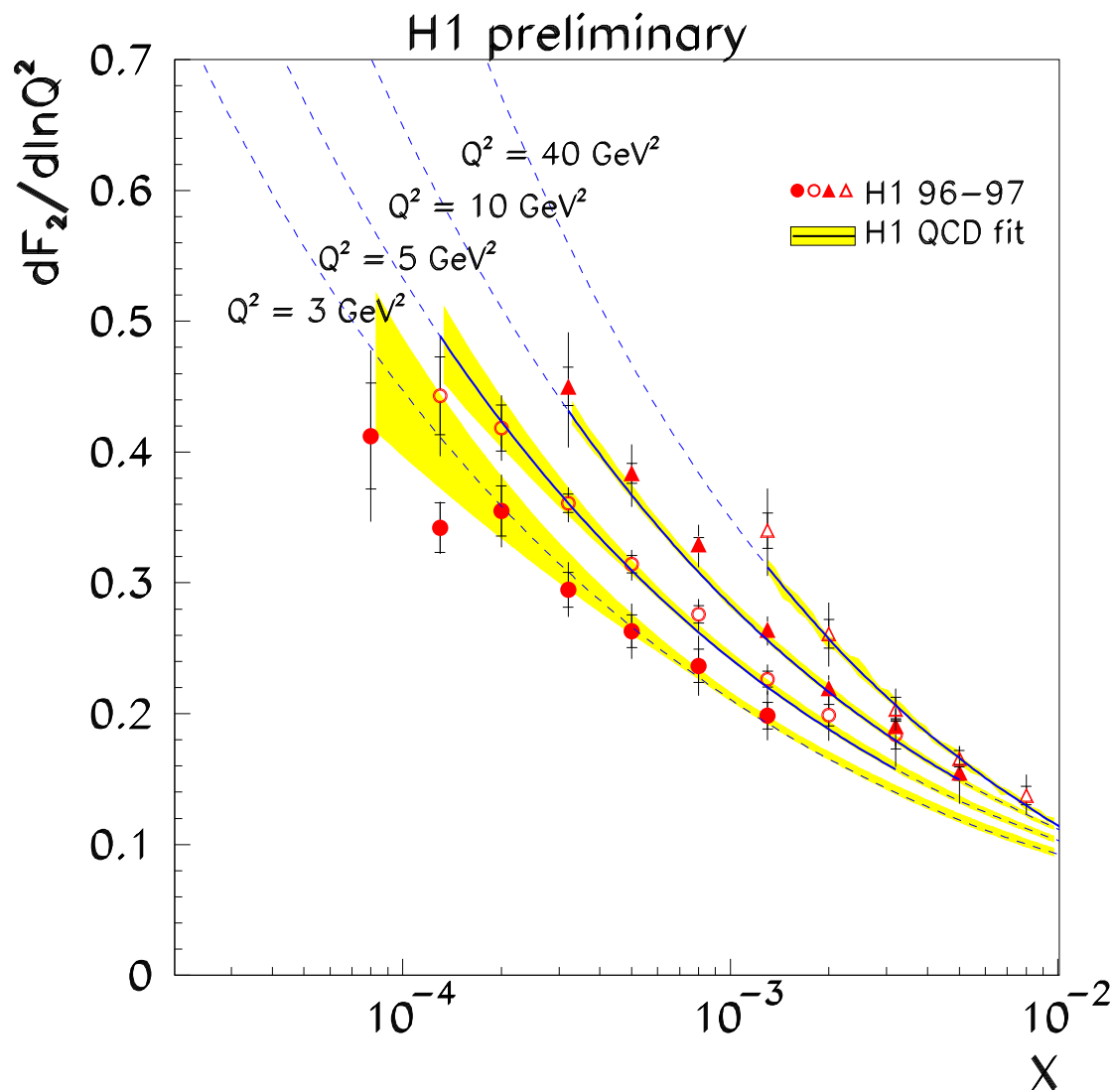}}
\end{center}
 \caption[junk]{The logarithmic slope of $F_2$, derived from H1 data, plotted as a function of $x$ at fixed values of $Q^2 > 3.5$~GeV$^2$. A monotonic increase of the slope with decreasing $x$ is observed at fixed $Q^2$. The dashed curves are extrapolations to regions outside of the data region considered in the fit.}
\label{fig:slopexh1} 
\end{figure}

\begin{figure}[htb] 
\vspace{2mm}
\begin{center}
\leavevmode
\hbox{%
\epsfxsize=5in
\epsfysize=7in
\epsffile{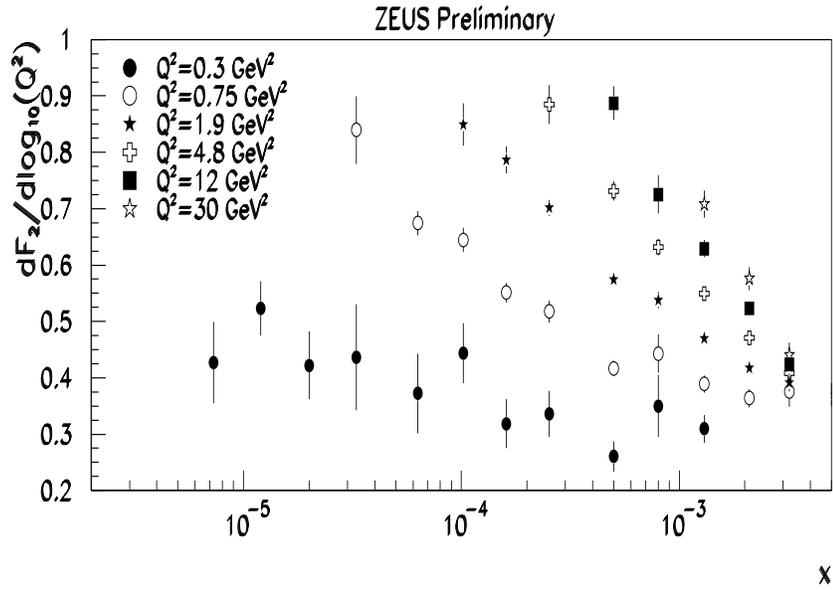}}
\end{center}
 \caption[junk]{The logarithmic slope of $F_2$ derived from ZEUS data plotted as a function of $x$ at fixed values of $Q^2$. A monotonic increase of the slope with decreasing $x$ is observed at fixed $Q^2$.}
\label{fig:slopexz} 
\end{figure}

To illustrate this transition region, the slope $\partial F_2(x,Q^2) / \partial \ln (Q^{2})$ 
as a function $x$ is often plotted in the low $Q^2$ region, 
to search for deviations from the logarithmic scaling violations predicted by perturbative QCD. 
This derived quantity is sensitive to the gluon density at small $x$. 
Figs.(\ref{fig:slopexh1},\ref{fig:slopexz}) show the latest plots of the slope derived from the  H1 and ZEUS data at fixed values of $Q^2$ 
(see e.g. \cite{h1osaka, yoshida, caldwell2} for more details). 
In the measured range, the slope is unambiguously seen to increase monotonically as $x$ decreases  even for $Q^2$ as low as $0.75$ ~GeV$^2$.  
An earlier, cruder version of this plot appeared to show a turnover in $x$ \cite{caldwell} and created a great deal of interest and discussion. It now appears from Figs.(\ref{fig:slopexh1}, \ref{fig:slopexz}) that this was a result of the limited range in $Q^2$ accessible at a particular $x$, and a changing range at different $x$ values, in the older data. 

\begin{figure}[htb] 
\vspace{2mm}
\begin{center}
\leavevmode
\hbox{%
\epsfxsize=5in
\epsfysize=7in
\epsffile{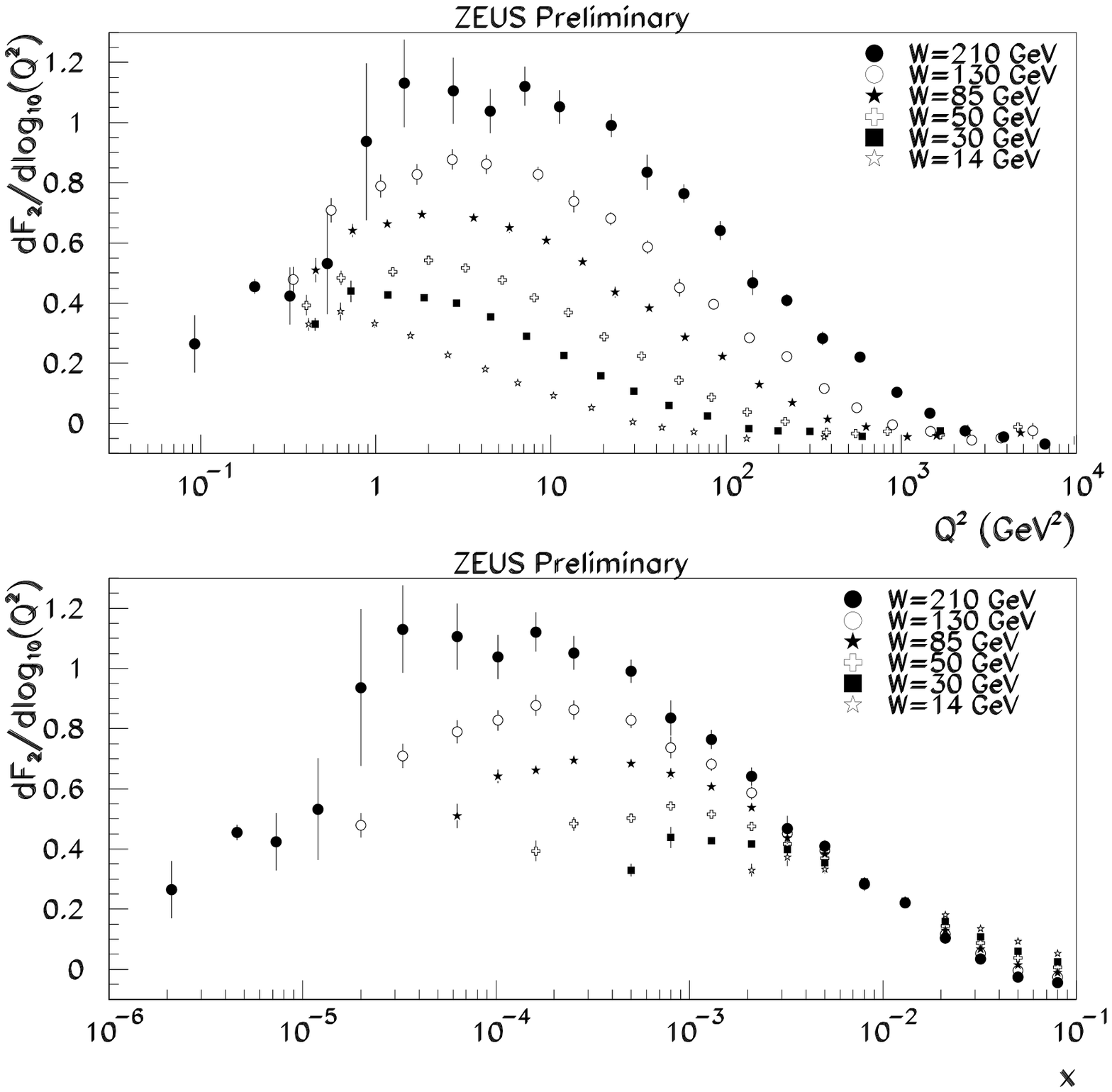}}
\end{center}
 \caption[junk]{The logarithmic slope of $F_2$ derived from ZEUS data plotted as a function of $Q^2$ (upper plot) and $x$ (lower plot) for fixed values of $W$. Note the apparent turnover in both plots for fixed $W$.}
\label{fig:slopew} 
\end{figure}

The lower plot of Fig.(\ref{fig:slopew}) \cite{caldwell2} shows the slope for fixed $W$, rather than $Q^2$. A genuine turnover in $x$ is revealed when the data is presented in this way. The upper plot shows the slope plotted as a function of $Q^2$, again revealing a turnover for fixed $W$. It is interesting to note that the position of the maximum appears to shift to higher $Q^2$ as $W$ increases.

The small-$x$ limit, which is also probed at HERA, corresponds to the 
high energy, or Regge, limit of DIS ($W^2 \gg Q^2, m_p^2,...$). 
The convergence of QCD perturbation theory in $\alpha_s(Q^2)$ is known to 
break down eventually here due to logarithms in $1/x$ 
(see e.g. \cite{forshaw} and references therein). 
Although the standard global DGLAP QCD fits produce an adequate description 
of the inclusive $F_2(x,Q^2)$ data, over a wide range of $(x,Q^2)$, certain 
characteristic features of the resulting fitted parton densities at 
low $Q^2$ and low $x$ suggest that this approach may require significant 
corrections. For example, the DGLAP analysis translates the 
experimentally-observed steep rise of $F_2$ at small $x$ into a numerically 
large gluon density, which rises steeply with $1/x$ 
(cf. Figs.(\ref{fig:slopexh1}, \ref{fig:slopexz})). 
Clearly, if parton densities get too large the partons must start to interact with one another in order to tame this growth (see e.g. \cite{glr}), 
leading to a breakdown of the standard DGLAP approach. 
This corresponds to the high density but weak coupling limit of perturbative QCD.

In addition, a startling experimental observation at HERA concerning 
DIS in the small-$x$ region is that an approximately constant fraction 
(of the order of $10 \%$) of events for $x \lsim 10^{-2}$, are diffractive 
(i.e. the proton remains intact rather than scattering inelastically). 
This is rather difficult to explain from within the standard parton model 
picture of a `typical' DIS event: in a `fast proton' frame, 
a highly virtual photon strikes a charged coloured quark, 
carrying collinear momentum fraction $x$ of the proton. This violent collision 
generates strong colour forces that rip the proton bound state apart. 
In this scenario, large rapidity gaps in the final state,  
which characterize diffractive events experimentally, are expected to be 
exponentially suppressed.
 
In the past few years the theoretical description of small-$x$ DIS has shifted 
somewhat from this `infinite momentum frame' to the frame in which the proton 
is at rest. This is particularly useful for considering diffractive scattering
(and especially the subclass of diffraction in which vector mesons and real photons 
are produced exclusively). In this frame, in the high energy limit, 
there is an appealing factorization of small-$x$ processes 
which in some sense is purely kinematic: 
the virtual photon fluctuates into a hadronic system at large distances  
upstream of the proton target, this hadronic state takes a long time to form,
interacts briefly with the target and subsequently forms the hadronic 
final state over a long period. From this frame it is much easier to 
believe that if the interaction is ``gentle''  
the integrity of the proton bound state will not be compromised.

The simplest fluctuation, which dominates for very small transverse 
size systems is the quark anti-quark state which forms a colour-triplet dipole. 
Because of this, the generic name of ``the dipole picture'' 
or ``the dipole approach'' has been adopted for proton rest frame 
descriptions. For larger transverse-size systems more complicated 
fluctuations, containing more than two partons, must also be considered. 
However, even for such complicated systems the transverse diameter of 
the fluctuation impinging on the target remains a useful variable to 
classify interaction strengths of the fluctuations of the virtual photon. 
For linguistic convenience, the transverse diameter and the interaction strength have 
become synonymous with the misnomers {\it dipole size} and {\it dipole cross section} 
respectively, regardless of the fact that more than two partons are typically involved for 
larger transverse-size fluctuations.

Within this framework, inclusive cross sections, for example $\sigma_L$, 
may be written in the following, generic, factorized way \cite{fs88,mue,nnn}:
\begin{equation}
\sigma_L (x,Q^2) =  \, \int_0^{1} dz \int d^2 \dperp \, 
{\hat \sigma} (\dperp^2,...) \, |\psi_{\gamma,L} (z,\dperp)|^2 \, .
\label{fl}
\end{equation}
\noindent In this simple case, the {\it light-cone wavefunction} of 
the longitudinally polarized virtual photon, in terms of the $q {\bar q}$ is given by QED:
\begin{equation}
|\psi_{\gamma,L}(z,\dperp)|^{2} = \frac{6}{\pi^{2}}\alpha_{e.m.} 
\sum_{q=1}^{n_{f}}e_{q}^{2}Q^{2}z^{2}(1-z)^{2} K_{0}^{2}(\epsilon \dperp) \, ,
\label{eqpsil}
\end{equation}
\noindent where $\epsilon^2 = Q^2 (z(1-z)) + m_q^2$ and 
$z$ is the fraction of the photon's momentum carried by the quark 
(of mass $m_q$). The variable $\dperp$ is Fourier conjugate 
to the relative transverse momentum of the $q {\bar q}$-pair and
in general corresponds to the transverse size (diameter) of the scattering system.
An appealing feature of this representation is that the short and long distance 
contributions to a given process are manifest. 
The power of the approach lies in the possibility that the interaction 
strength depends only on the configuration of the interacting system and not on 
how it was produced. For example the strength depends on the system's transverse size 
and not on external variables such as $Q^2$. 
This implies a {\it universal} interaction cross section, ${\hat \sigma}$, 
which may be used in a wide variety of small-$x$ inclusive, exclusive and 
diffractive processes. 

There are several models available for this interaction cross section 
${\hat \sigma}$ and we compared and contrasted four of them \cite{mfgs,glm,wgb,kfs} 
in detail at the meeting. A summary of these discussions is given in the Appendix. 
One issue is precisely how this  cross section depends on the energy of 
the interaction, i.e. whether it depends on $W^2$ or $x$ (bearing in mind that it must reproduce 
the observed approximate scaling properties of the structure function $F_2 (x,Q^2)$). 
In principle, ${\hat \sigma}$ can also depend on $z$, although all models constructed so far neglect this dependence.
There are many approaches influenced by similar ideas covering various aspects of small-$x$ physics. 
In this short note we make no attempt to be complete in covering all references and 
instead refer the reader to the references within \cite{mfgs,glm,wgb,kfs} 
and to recent review articles on diffraction \cite{heb,mw,ht}.

Broadly speaking, each of the models discussed share some common features designed 
to encapsulate the important physics issues. For small $\dperp$ the dipole cross section 
is ascribed a steep rise with energy (from the point of view of QCD this  
corresponds to the steeply rising gluon density). A much more gentle (or even flat) 
behaviour in energy is ascribed for larger non-perturbative sizes (in agreement with observations 
of hadronic cross-section data). There are two important consequences of this.
Firstly, each model contains a mixture of perturbative and non-perturbative  
aspects. The interplay between them may be systematically studied,  
within a particular model, by examining the $\dperp$-integral for the process 
of interest. Generally, the hard perturbative part becomes more important as 
the energy increases (since it grows faster than the soft piece). Secondly, 
the cross sections associated with small dipoles reach hadronic sizes (tens 
of millibarns) in the kinematic region probed at HERA 
(because the gluon density is numerically large at small $x$). 
The latter feature implies that some kind of ``taming'' or saturation 
physics is necessary at high energies in order to preserve the 
natural geometrical criterion that the dipole cross section increases 
monotonically with transverse size. The introduction of such taming effects 
implies a new high-density regime of hard QCD interactions.

Our primary goal is to identify clear signs for taming effects in 
experimental measurements. Of course, in order to claim that we have 
found something new, we need to understand more precisely how the 
dipole picture relates to the more rigorous DGLAP formalism 
in the kinematic region where they are supposed to agree (moderate $Q^2$ 
and small, but not too small, $x$). In practice, since the analysis of 
hadronic interactions is always complicated by our 
poor understanding of confinement, it will probably be necessary to build 
up a body of experimental evidence, combining data from several 
experimental channels, in order to force the unambiguous conclusion that we 
have entered a new regime. 

The connection of the dipole approach with standard DGLAP is 
considered in Section 2. The issue of `higher twist' corrections to DGLAP 
is discussed briefly in Section 3. Section 4 covers the description of 
inclusive diffraction and exclusive diffractive processes (such as 
vector meson production) within the dipole framework. The connection 
to the DGLAP evolution of diffractive parton densities is also briefly discussed. 
Finally, in Section 5 we conclude with a `wish list' for better measurements 
of particular experimental processes that will pin down the details of 
the dipole cross section and may reveal saturation phenomena, thus 
achieving a principle aim of the meeting, i.e. to help prioritize the 
experimental effort in this area.

We note in passing that although we were mainly concerned with high energy 
interactions of photons and protons, the concept of a universal high energy 
interaction cross section, embodied in ${\hat \sigma}$, is in principle 
more widely applicable. It may be used for example for DIS using 
nuclear targets, in photon-photon collisions 
and even for certain hard processes in hadron-hadron interactions.

\section{Comparison of the dipole picture with the standard DGLAP analysis}

Clearly the dipole approach to small-$x$ processes should be complementary to the 
DGLAP analysis (for moderate $Q^2$ and not too small $x$). 
In this section we briefly recap the standard approach and then 
emphasize similarities and differences between the two pictures.

The standard perturbative QCD approach to calculating the structure functions of DIS  
invokes collinear factorization and involves a convolution of universal long-distance 
parton (quark and gluon) density functions (``pdfs'', into which all infrared divergences are factorized)
specified at the factorization scale, $\mu_f^2$, with short-distance, hard coefficient functions 
for the fusion of these partons with the virtual photon. The coefficient functions are calculable in 
perturbative QCD as a power series expansion in $\alpha_s$. 
This power series generates logarithms in $Q^2/\mu_f^2$ 
to all orders in perturbation theory (usually one chooses $\mu_f^2 = Q^2$ 
to suppress these potentially large logs in the coefficient functions and ``move them'' 
to the pdfs). The independence of physical cross sections on the choice 
of this factorization scale then leads to the DGLAP equations \cite{dglap}, 
which embody the physics of the renormalization group and describe 
how the pdfs change as their scale changes. The pdfs, which are solutions to
the DGLAP equations, are then specified to a given order in logarithms (LO, NLO, NNLO\footnote{For recent progress on NNLO calculations for parton densities and structure functions see e.g \cite{vnvogt}.},...), with 
the accuracy increasing as more and more terms are included in the perturbative expansion.
The parton densities at the input scale are not calculable in QCD, but factorization 
implies that they are {\it universal} functions. In practice, the various groups responsible 
for producing pdfs (CTEQ, MRS, GRV) constrain their analytic forms in $x$ at the input scale, 
$Q_0^2$, using sum rules, symmetries, etc. The non-perturbative parameters used to specify 
the input distributions are then determined by performing a minimum $\chi^2$ fit to all 
available relevant data, evolving to the appropriate scale using leading-twist DGLAP evolution. 
The value to choose for the input scale, $Q_0^2$, is not predicted by the theory. 
Recently scales as low as $Q_0^2 = 1.0$ ~GeV${^2}$, or even lower in the case of GRV \cite{grv98}, 
have been used.

The factorization is termed collinear because all transverse momenta 
(defined relative to the photon-proton axis) in the pdfs is forced to be less than $\mu_f^2$. 
As far as the photon is concerned it strikes a parton which has negligible transverse 
momentum relative to the proton's direction, so all input partons are moving collinear 
with the proton.

{}From the theoretical point of view this collinear factorization, which separates  
short and long distance contributions, is made possible by the use of the operator 
product expansion (OPE). For very large $Q^2$, a subclass of all 
possible Feynman graphs are dominant in which, for a particular choice of gauge, the 
hard coefficient function is connected to the proton by only two parton lines (the 
so-called ``handbag graphs'' and their perturbative QCD corrections). 
Clearly, for sub-asymptotic $Q^2$ this picture has corrections which the OPE predicts 
contribute at relative order ${\cal O} (1/Q^2)$ and beyond (${\cal O} (1/Q^2)^n, n=2,3,...$). 
These are commonly called ``higher-twist corrections'', although it should be remembered that 
within the OPE twist has a rigorous definition\footnote{For a given operator the twist is the difference between its mass dimension and spin, and for a given product of operators it 
is the sum of the twists of the individual operators.}. 
For example, a subclass of the most important sub-asymptotic corrections (twist four) 
corresponds to the set of Feynman graphs in which the hard coefficient function 
is connected via four parton lines to the proton.

For very small $x$ the standard implementation of DGLAP must break down due to the development 
of logarithms in energy or $1/x$ at each order in perturbation theory, which eventually must spoil 
the convergence of the perturbative series because the product $\alpha_s (Q^2) \ln (1/x)$ 
becomes ${\cal O}(1)$.  The leading logarithms in $1/x$ are 
re-summed by the BFKL equation \cite{bfkl} and recently next-to-leading 
logarithmic corrections to the leading-log result were completed 
\cite{nlobfkl} and were found to constitute a very large correction to the 
leading-log result. Subsequently, considerable progress has been made 
in understanding the reasons for the large size of these corrections 
(for a recent overview and references see \cite{salam}).

A long standing objection to the use of BFKL calculations was again 
raised by L. Frankfurt in the discussion session, i.e. that the HERA 
kinematics provide a relatively restricted rapidity range 
for perturbative real gluon emission in quasi-multi-Regge kinematics. 
Realistically, HERA kinematics allow for typically two or three real gluon 
emissions on average, so that summing an infinite tower of gluon ladders 
may be unrealistic. Moreover, the application of these re-summed calculations 
to relevant physical processes are plagued by many uncertainties, including 
in many practical situations an uncontrollable contamination of 
non-perturbative physics. The final solution to this extremely difficult 
problem is far from clear. This fact in itself lends support to, and makes 
timely, phenomenological frameworks such as the dipole picture which embody 
a certain amount of modelling concerning the mixing of perturbative and 
non-perturbative physics. The interplay of new experimental 
results and the refinement of these models then becomes of paramount 
importance in attempting to pin down the relevant physics.

For  small $x$ it is necessary to generalize collinear factorization 
to what is known as ``$k_{\perp}$-factorization'' \cite{catani} and to discuss unintegrated 
distributions rather than conventional pdfs (the subject of the BFKL equation is the 
unintegrated gluon density). At leading-twist level, the former are related to the latter 
via an integration over transverse momentum, e.g. for gluons

\begin{equation}
xg(x,Q^2) = \int^{Q^2}_{0} \frac{d \lperp^2}{\lperp^2} f (x,\lperp^2) \, .
\end{equation}

\noindent The connection between the dipole picture and standard DGLAP (and especially 
$k_{\perp}$-factorization) is far from obvious. Clearly both pictures involve a 
photon connected to a quark loop. The transverse momentum and energy fractions used to 
specify the $\psi(z,k_{\perp})$s in the dipole approach are closely related to the 
momentum flowing around this loop. It would be extremely interesting and useful 
from a theoretical point of view to understand collinear factorization in $\dperp$-space 
more precisely. 

One recent model \cite{mfgs} claims to include as much information as possible from DGLAP 
in that it is explicitly built using gluon densities from DGLAP fits 
(implicitly incorporating DGLAP evolution) and the well-known leading $\ln Q^2$ formula for 
the dipole cross section:
\begin{equation}
{\hat \sigma_{pQCD}} (x,\dperp^2) = \frac{\pi^2 \dperp^2}{3} \alpha_s ({\bar Q}^2) 
xg(x,{\bar Q}^2) \, , 
\label{eqpqcd}
\end{equation}
which has been widely used in the analysis of various diffractive channels 
(for a review see \cite{mw}).
Indeed, two-gluon exchange has been explicitly proven to be the dominant contribution in certain 
hard exclusive processes \cite{cfs,cf}. The eikonal model of \cite{glm} also reduces to precisely  
this form for the limit of very small $\dperp$ (see Appendix). 

Unfortunately, an additional assumption concerning the connection between transverse sizes and 
four-momentum scales is unavoidable. This leads to a specific model-dependent assumption 
about the connection between the dipole picture and standard DGLAP. 
Having made this connection (using ${\bar Q^2} = 10/\dperp^2$) the model of \cite{mfgs} 
uses eq.(\ref{eqpqcd}) over as wide a range of transverse sizes as possible, i.e. 
for all $\dperp < d_{\perp,Q0} $ (where $d_{\perp,Q0} = \sqrt{10} / Q_{0}$  
corresponds to the input scale $Q_0$). The model of \cite{glm} makes the choice 
$d_{\perp,Q_0} = 2/Q_{0}$ instead. Of course, both choices are equivalent to 
leading-log accuracy.

The nature of $Q^2$ evolution in the dipole picture is particularly interesting.
At a particular $Q^2$, the integrand of a given cross section, such as $\sigma_L$ of 
eq.(\ref{fl}), has a profile in $\dperp$. In this sense, the $\dperp$-profile for $F_2$ at 
$ Q^2 = Q_0^2 $ ``corresponds'' to a model for the input quark density. 
The light-cone wavefunction of the photon clearly changes as $Q^2$ increases and 
this effect ``evolves'' the profile to smaller $\dperp$. This also means that the gluon density
in eq.(\ref{eqpqcd}) is predominantly being sampled at typically larger ${\bar Q}^2$ scales. 
This shift in the profile in $\dperp$ with $Q^2$ is very closely related to standard DGLAP 
evolution but unfortunately, a more precise connection can not be made at present.

The other two models \cite{wgb,kfs} were ``designed'' to describe the transition 
from the photoproduction region to DIS (from small to moderate $Q^2$). Hence, the known QCD behaviour 
at small $x$ (steeply rising gluon densities) was included in a more phenomenological 
way by including a piece that rises steeply with energy at small $\dperp$ (see also \cite{capella}).

Overall at the meeting there was a broad consensus that eq.(\ref{eqpqcd}) imposes the correct QCD behaviour, to leading-log accuracy, for small dipole sizes.

\section{Higher twists}

%{\it Do Jochen Bartels or Genya Levin want to add more ? I'd particularly like to include
%Genya's probabilistic leading log $Q^2$ statement about twist-four graphs, if it can 
%be put succinctly and clearly. Perhaps it would also be good to show a diagram, e.g. Fig.(1c) of Jochen's paper with Krzysztof and K. Peters.  I am not knowledgeable enough to say more than the following:}

Unfortunately, from a theoretical point of view, very little can be said about the numerical importance of higher twist effects, at a given $Q^2$, with a high degree of rigour.

One of the most basic and perhaps one of the more important higher twist corrections 
involves graphs where four gluons couple directly from the proton to the quark box. 
An analysis \cite{bgbp} of such graphs revealed an intriguing partial cancellation 
between the correction to the longitudinal structure function, $F_L$, and the transverse one, $F_T$, when they are summed in $F_2$ 
(see also \cite{bartels} for an analysis of related graphs). 
This cancellation has been taken by some supporters of the standard approach 
as an explanation as to why the leading-twist DGLAP analysis is so successful. 
In addition, this cancellation was also seen explicitly within the context of a specific 
model \cite{wgb} for ${\hat \sigma}$. If the cancellation is genuine in $F_2$ it provides 
yet another strong motivation to make an explicit experimental measurement of $F_L$ 
in the region of moderate $Q^2$ (for example by lowering either the proton or lepton 
beam energy at HERA) to look for deviations from the leading-twist predictions.
Of course, in order to be discriminatory, this would require a sufficiently long running period 
to make a high precision measurement of $F_L$.

During the discussion sessions 
the point was made by L. Frankfurt that conclusions based on specific 
subclasses of twist-four graphs may be too naive. He 
speculated that perhaps once twist-four corrections 
become numerically important all other higher twists may as well 
(i.e. the twist expansion could break down). 
This scenario would imply that the system undergoes some kind of phase transition to a completely new regime.

\section{Diffraction and heavy flavours}

\subsection{Exclusive processes as probes of ${\hat \sigma}$}

\begin{figure}[htbp]
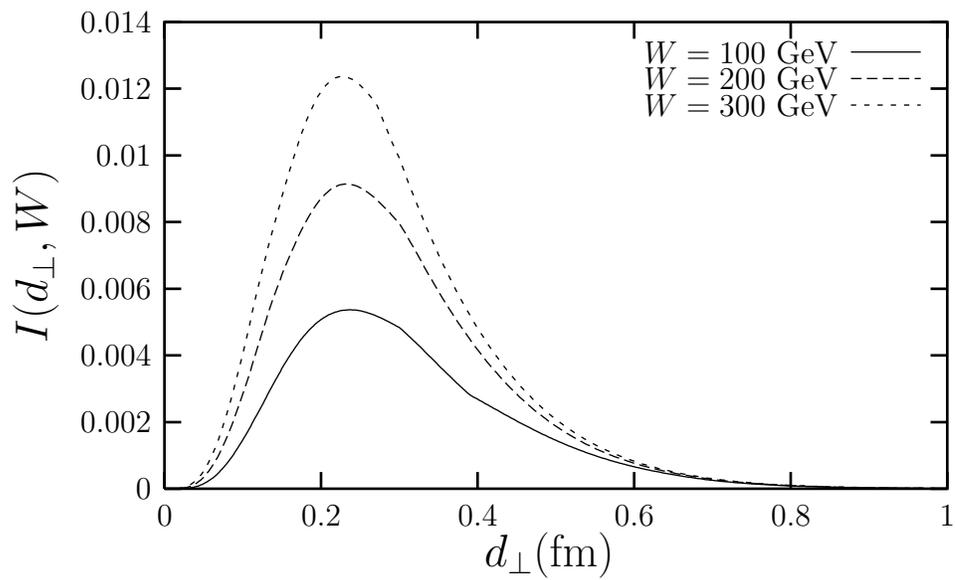

  \begin{center}%
	\include{djpsiint}
\caption{The profile in $d_{\perp}$ for the photoproduction of $J/\psi$, using the dipole model of \cite{mfgs}, at three different energies.}
    \label{sigdint}
  \end{center}
\end{figure}

In hard exclusive processes, such as diffractive vector meson production and deeply virtual Compton scattering, ${\hat \sigma}$ enters on the 
amplitude level, rather than on the cross-section level. For example, the differential cross section
for vector meson production is given by:
\begin{eqnarray}
\frac{d \sigma }{d t} \left|_{t=0} \right. &\propto& |A(s,t=0)|^2 \, , \\
A & = & \int d^2 \dperp ~d z ~\psi_{\gamma} (z,\dperp) ~{\hat \sigma} ~\psi_V (z,\dperp) \, ,
\label{eqvm}
\end{eqnarray}
\noindent where $\psi_V (z,k_{\perp})$ is the light-cone wavefunction of the produced vector meson. 
Thus, these exclusive processes are sensitive to integrals that weight ${\hat \sigma}$ differently in 
$\dperp$ than inclusive structure functions.  As an example,  
Fig.(\ref{sigdint}) shows the integrand, or profile, in $\dperp$ of the amplitude for photoproduction 
of $J/\psi$ at three different values of $W$ in the HERA range. A hybrid light-wavefunction for $J/\psi$ from \cite{fks} 
was used and the model of \cite{mfgs} for ${\hat \sigma}$. Moving to  larger $Q^2$ would  
cause the peak and profile in $\dperp$ to shift to small distances as the process becomes 
more perturbative. Varying $Q^2$ and $W^2$ and looking at different states 
($\gamma, \rho, \phi, J/\psi, \Upsilon$) allows a detailed scanning of ${\hat \sigma}$ over a wide range in $\dperp$ and energy. Therefore, in principle exclusive processes provide 
complementary information to inclusive structure functions that may be used to pin down the 
detailed functional form of the dipole cross section.

Unfortunately, these processes are sensitive to several other uncertainties, 
including the effects of skewedness of the amplitude, model dependence associated 
with $\psi_V (z,k_{\perp})$, poorly known $t$-slope, etc, which make uncovering the details of 
${\hat \sigma}$ a difficult job in practice. To obtain the maximum information, a
simultaneous analysis of all available data within the same framework is required.

\subsection{Inclusive diffraction}

One of the appealing features of the dipole picture is its ability to describe diffractive 
and inclusive scattering within the same framework. The $q \bar q$ dipole produced by a 
longitudinally-polarized photon has the following contribution to inclusive diffraction:
\begin{eqnarray}
\frac{d \sigma }{d t} \left|_{t=0} \right. &=& \frac{1}{16 \pi} \int d^2 \dperp dz 
|\psi_{\gamma,L} (z,\dperp, Q^2)|^2 |{\hat \sigma}|^2 \, .
\end{eqnarray}
\noindent Note that ${\hat \sigma }$  appears squared. Roughly speaking, one may think of this 
as the square of the amplitude in eq.(\ref{eqvm}), but instead of an explicit vector meson 
wavefunction appearing completeness in the sum over all possible produced states is used.

The proof by Collins of factorization in hard diffractive scattering \cite{collins}
placed the DGLAP analyses of diffraction in DIS on a firmer theoretical footing. 
The fundamental objects which evolve with $Q^2$ are known as diffractive parton densities. 
Relative to standard pdfs, they are defined with the additional constraint that the proton 
remains intact in the final state. As such they depend on two extra variables which parameterize 
the scattering proton ($\xp $, the longitudinal momentum fraction lost by 
the incoming proton, and $t$, the square of the four-momentum transfered to it, are usually chosen). 
With the additional assumption of Regge or Ingleman-Schlein factorization, a 
DGLAP analysis \cite{actw, h1diff} leads to the conclusion that the diffractive gluon density 
is much larger than the diffractive quark density for values of 
$\beta = x / \xp \gsim 0.1$  (the ratio of gluons to 
quarks is several times larger in diffraction than in the inclusive case).

In diffraction, the connection between the dipole picture and DGLAP-driven diffractive 
parton densities is more obscure. Qualitative statements, similar to those made in the 
inclusive case, concerning the evolution of the $\dperp$-profile as $Q^2$ changes may 
also be made here. A further essential complication is that one must 
also include higher order Fock states (including $q {\bar q} g$) in the virtual photon,  
especially for the case of large diffractive masses\footnote{In contrast, inclusive structure functions are unique in that it is possible to factorize all gluon emission into the definition of the dipole cross section. For hard vector meson production the $q {\bar q} g$ contributions are suppressed by the light-cone wavefunction of the vector meson.}. The ratio of diffractive to inclusive 
total cross sections is particularly interesting since it also helps to specify the dominant $\dperp$.
The fact that this ratio appears to be fairly independent of energy supports the 
widely-held view that inclusive diffractive processes are much more sensitive to 
larger transverse sizes.

\subsection{Diffractive heavy flavour and jet production}

Diffraction can also occur for processes involving a colour-octet pseudo-dipole in the initial state. 
These result from asymmetric $q {\bar q} g$ fluctuations in which the quark and anti-quark 
remain close together in transverse space. This may be achieved by demanding heavy flavours or 
a large relative $p_{\perp}$ between the quark and anti-quark, with the gluon carrying a 
small fraction of the available photon energy.
Such fluctuations are closely related to diffractive gluon densities (see \cite{heb} for 
more details) and have become known as the ``gluon channel'' for diffraction.
These configurations are known to have a large colour factor (9/4) relative to the 
colour-triplet dipole (the ``quark channel''). 
This large factor may make the overall $\hat \sigma$ for hard gluon-channel diffractive processes 
much closer to the saturation limit. As such, it should provide a clear signal for new 
physical effects in the saturation regime (see \cite{fs} for a longer discussion of 
these issues). The difficult part is finding a physical process in which one is sensitive 
to small enough colour-octet dipoles to observe taming effects in a perturbative environment.

If the large colour factor is translated directly to ${\hat \sigma}$ it leads to some 
striking predictions for diffractive channels that are driven by the exchange of gluons 
and dominated by colour-octet fluctuations of the photon. 
One excellent channel to consider is diffractive heavy flavour production. 
The above observation, concerning the large ratio of diffractive gluon density to diffractive 
quark density relative to its inclusive counterpart, naively suggests the striking 
prediction that about $30 - 40 \%$ of charm quark production should be diffractive. 
Realistic kinematic constraints, including a typical effective scale of about $10$ ~GeV$^2$,  
implemented in the RAPGAP Monte-Carlo \cite{rapgap}, suggest that in practice one may expect a ratio 
closer to $20 \%$. 
This effect is not yet seen in the currently measured region, which is concentrated at 
rather low values of $\beta \approx Q^2 / (Q^2 + M_X^2)$ because of the necessity to 
produce the heavy charm quark pair (according to ZEUS \cite{zosaka} the diffractive charm 
appears to be about $6\%$ of the total). With the expected higher luminosity it will be possible 
to collect more data at higher values of $Q^2$ and hence $\beta$. An observation of a 
sharp increase in the diffractive charm ratio would be a strong indication of the 
correctness of this picture.

Another gluon-driven channel is the diffractive production of jets (both two and three jet events 
are interesting). One experimental problem here is that the requirement to have a 
well-defined jet naturally forces one to consider a large $M_X$ system and hence one is 
pushed again into the low $\beta$ region (high $\xp$  at fixed $x$). 
Again, the higher luminosity will allow more data to be collected at higher $\beta$.

\section{Conclusions and a wish list of desirable measurements}

The H1 and ZEUS experiments have each collected over 100 pb$^{-1}$ of data so far. With the 
increase of luminosity after the HERA upgrade comes the opportunity to  
improve considerably the precision of a large number of measurements of diffractive processes. 
At this meeting, we tried to prioritize which of these processes are the most important to 
measure well in order to determine whether a new saturation regime really has been reached 
at small $x$.

{}From the point of view of the dipole picture we need better measurements to pin down 
the energy and transverse size dependence of the universal function ${\hat \sigma}$. 
As discussed above, this may be realistically achieved only by a combined analysis of 
several exclusive and inclusive processes.

%MM Pat requests more explicit here
%> - Your shopping list in 2nd paragraph of Section 5 needs to be made more
%>   explicit. The experiments have been doing a fairly exhaustive job of  
%>   studying the vector mesons, for example. WHat exactly do you wish to see
%>   done?

The precision measurements of $F_2$ are now excellent and very high precision data exist 
over a very wide kinematic range. Indeed, the main uncertainties in $F_2$ over a wide 
range in $Q^2$ are now given by systematic rather than statistical errors. It would therefore 
be extremely useful to have the complimentary information that a dedicated 
high-precision measurement of $F_L(x,Q^2)$ in the same kinematic range would provide. 
This would help to constrain the dependence of ${\hat \sigma}$ on $\dperp$ and with 
sufficient precision could act as a test of the DGLAP parton densities. 
A serious deviation from the DGLAP predictions (determined largely from the measurement of $F_2$) 
may confirm the significance of higher twists in $F_L$ (see section 3). 

Improved statistical precision on the derived quantity $\partial F_2 / \partial \ln Q^2$ 
in the region of relatively low $Q^2$ would be a good 
discriminator between approaches and a combined quantitative analysis of this information 
with the description of $J/\psi$ production within the same framework was advocated by U. Maor. 
He claimed that none of the currently available pdfs were able to fit 
both quantities simultaneously, without including taming effects \footnote{Since the meeting 
a preprint \cite{gflmn} has appeared.}. Of particular interest is further constraining the energy 
dependence of hard (either high $Q^2$ or large mass $M_V^2$)  vector meson production 
in the highest possible bins in $W^2$. An observation of a flattening of the steep rise 
with energy would constitute a clear and relatively model-independent signal for taming 
in these processes.

Improved measurement of diffractive charm and diffractive dijets, especially at higher values of 
$Q^2$ would be particularly welcome since they are sensitive to gluon induced processes in 
which saturation effects may be seen early.

There was a consensus of opinion that it was vital that the $t-$dependence of various diffractive processes should be measured, including, if possible, how it changes with energy (shrinkage). 
The $t$-dependence is usually parameterized in exponential form, $e^{(B_{D} t)}$, 
using the $t$-slope parameter, $B_D (W^2)$. The rate of decrease of  
this diffractive peak with increasing $t$ and $W^2$ for a given diffractive process, 
embodied in $B_D$, gives a strong indication of the mixture of soft and hard contributions. It was agreed that the measurement of $B_D$ is of major importance.

From an experimental point of view, we are strongly encouraged by 
the proposal of H1 to insert a dedicated detector in the very forward region 
to detect scattered protons \cite{favart}. This detector will allow a clean 
tagging of diffractive interactions, taking full benefit of the high luminosity 
in the new HERA running regime. This is particularly important for measurements of the diffractive structure function (where the major uncertainty is associated with the presence of proton dissociative background), and of jet and of charm 
diffractive production. Most relevant for the present discussion will be the 
possibility of measuring the slope parameter, $B_D$. 

In conclusion, the first hints of a new saturation regime may already have been seen at HERA, 
in a rather indirect way, using the conventional framework of DGLAP.
The phenomenological dipole approach is capable of producing clear predictions 
for experimental signatures that would reveal this new regime and the relevant calculations 
are already underway. There are excellent prospects for improved measurements of many 
small-$x$ processes, which are capable of confirming this new regime directly. 
In practice, it appears to be necessary to perform a combined quantitative analysis of current 
and future data for several different channels to build up the necessary evidence.

\section*{Appendix: Features of different models for ${\hat \sigma}$}

The aim of this section is to compare and contrast several 
models \cite{mfgs,glm,wgb,kfs}  for the dipole cross section and 
how they depend on $\dperp$ and energy. This section is included mainly 
for information and no attempt is made to be critical of physics used 
in building particular models. 
A more critical discussion from the point of view of the author and 
his collaborators may be found in Section 4 of \cite{mfgs}. 
While much of the discussion 
in Amirim, perhaps inevitably, centred on the contrast between the different 
approaches, a tentative consensus of the discussion sessions was 
that the features which the various models share should eventually 
prove to be much more significant than the details in which they differ. 
The most important of these shared features was the implementation of some
kind of taming of the rapid growth of the short distance part of ${\hat \sigma}$ 
as the energy increased.

First of all we consider the simple saturation model for ${\hat \sigma}$ 
suggested by Golec-Biernat and W\"{u}sthoff \cite{wgb}:
\begin{equation}
{\hat \sigma} (x,\dperp^2) = \sigma_0 ( 1 - \exp [-\dperp^2 Q_0^2/ 4 (x/x_0)^\lambda ] ).
\label{eqwgb}
\end{equation}
\noindent A three parameter fit to the HERA data on $F_2$ with $x < 0.01$, 
excluding charm and assuming $Q_0 = 1.0$ GeV, produced the following values 
$\sigma_0 = 23.03 $~mb, $x_0 = 0.0003$, $\lambda = 0.288$ and a 
reasonable $\chi^2$. This model also did an adequate job in describing the 
diffractive data (see second reference of  \cite{wgb}) without any further tuning of parameters 
(but with some additional assumptions, for example that exactly the same 
${\hat \sigma}$  also couples, in a particular way, to the $q {\bar q} g$ component 
which is required to describe inclusive diffraction).

Certain features are apparent in eq.(\ref{eqwgb}).  First of all,  
${\hat \sigma}$ is assumed to depend only on the transverse size, $\dperp$, and upon 
the Bjorken scaling variable $x$. For small transverse size,  
$\dperp^2 \ll 4 (x/x_0)^\lambda /Q_0^2$, ${\hat \sigma}$ 
is designed to be proportional 
to the area of the scattering system 
(a feature expected on geometrical grounds) and to grow steeply with $1/x$ 
at small $x$:  ${\hat \sigma} (\mbox{small} ~\dperp) \propto \dperp^2 ~x^{-0.3}$. 
In contrast, at large $\dperp$ flat behaviour in $\dperp$ and $x$ is assumed: 
$\sigma (\mbox{large} ~\dperp) \rightarrow \sigma_0$. As $x$ decreases this flat 
region extends to smaller and smaller $\dperp$.

For certain exclusive electroproduction processes such as Vector Meson 
Production \cite{cfs} and Deeply Virtual Compton Scattering
\cite{cf} the interaction of a small $q {\bar q}$ has been rigorously 
proved to proceed predominately through the exchange of two gluons in 
the $t$-channel. Thus perturbative QCD relates ${\hat \sigma}$, to leading 
log($Q^2$) accuracy, to the leading-log gluon density in the proton 
(for a detailed derivation see \cite{rad}):
\begin{equation}
{\hat \sigma_{pQCD}} (x,\dperp^2) = \frac{\pi^2 \dperp^2}{3} \alpha_s ({\bar Q}^2) 
xg(x,{\bar Q}^2) \, .
\label{sigpqcd}
\end{equation}

\noindent To leading-log accuracy one may make the choice ${\bar Q}^2 = Q^2$. 

A detailed model for ${\hat \sigma}$ based on this form is given in $\cite{mfgs}$ 
for all transverse sizes. The detailed form relies on the continuity and 
monotonic increase of ${\hat \sigma}$ as a function of $\dperp$ (at fixed $x$). 
For large, non-perturbative, transverse distances ${\hat \sigma}$ is matched 
on to a typical hadronic cross section 
(pion-proton is used at $\dperp = d_{\perp, \pi} = 0.65$ fm). This is ascribed a 
slow Donnachie-Landshoff energy growth typical of hadronic cross sections
\footnote{In order to preserve the uniqueness of ${\hat \sigma}$ as a function of 
$x$ and $\dperp$, it was necessary to implement this energy growth via a 
multiplicative factor $(x/x_{0})^{0.08}$ rather than the usual $(s/s_{0})^{0.08}$ behaviour. 
Since $x_0$ and $s_0$ are in principle unknown, this is an acceptable 
compromise over a reasonable range in energy.}. 

The model in \cite{mfgs} attempted to go beyond leading-log accuracy 
in ${\hat \sigma}$ by incorporating a $\dperp$-dependence for the four-momentum scale 
${\bar Q}^2 = 10/\dperp^2$. In addition, momentum conservation is implemented 
for the fusion of the photon and the gluon. This leads to the gluon being sampled 
at an effective momentum fraction $x'$ which depends explicitly on the mass of 
the produced $q {\bar q}$-pair (which may be translated into a $\dperp$ 
dependence in the simplest case). With these modifications in place the 
form given in eq.(\ref{sigpqcd}) is used within its region of applicability, 
i.e. for $\dperp^2 < 10 / Q_{0}^2$, providing $x$ is not too small. 
Note that the generic geometric $\dperp^2$ behaviour becomes modified by the  
behaviour of $\alpha_s xg$ on ${\bar Q}^2$ which translates into a complicated 
effective behaviour on $\dperp$ (which will depend on the value of $x$). 

For very small $x$, it was found that using CTEQ4L leading-log partons 
the numerical value of the dipole cross section at small $\dperp$ became large 
(of hadronic interaction strength) due to the numerical size and steep 
growth of the gluons at small $x$. This observation threatens to spoil 
the monotonic increase of ${\hat \sigma}$ with $\dperp$ expected on geometrical grounds 
and was taken to be a signal for the necessity of taming corrections  
to the exchange process. These were implemented by hand in \cite{mfgs} 
at fixed $x$ at the critical point, $d_{\perp, crit} (x)$, at 
which ${\hat \sigma}$ reached $50 \%$ of its maximum value (as determined by the pion-proton 
cross section). The perturbative QCD expression for ${\hat \sigma}$ in 
eq.(\ref{eqpqcd}) then has a progressively reducing region of applicability 
as $x$ decreases: $\dperp < d_{\perp, crit} (x)$ (cf. the ``critical line'' of \cite{wgb} and the packing factor of \cite{glm}).

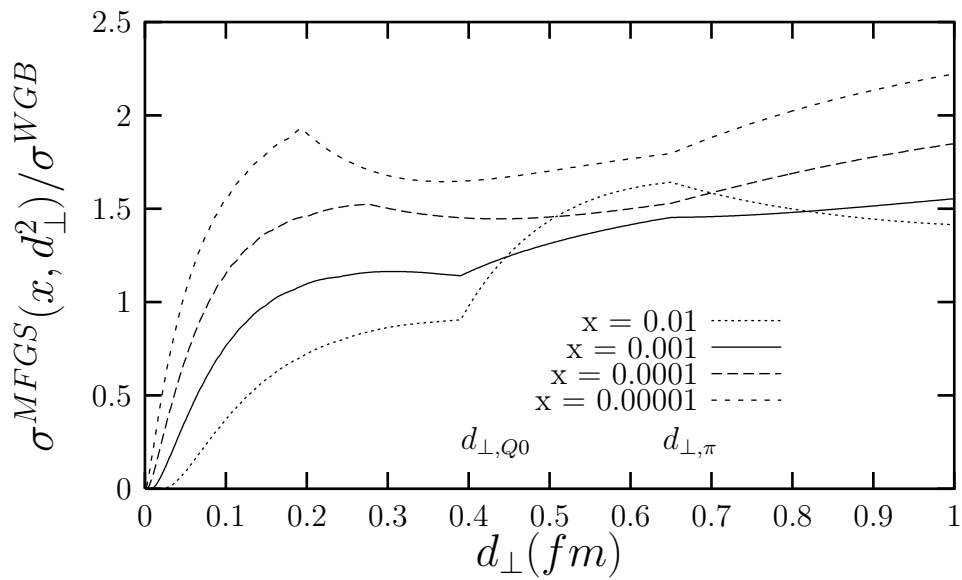
\begin{figure}[htbp]
  \begin{center}
     % GNUPLOT: LaTeX picture with Postscript
\begingroup%
  \makeatletter%
  \newcommand{\GNUPLOTspecial}{%
    \@sanitize\catcode`\%=14\relax\special}%
  \setlength{\unitlength}{0.1bp}%
\begin{picture}(3600,2160)(0,0)%
\special{psfile=wgb llx=0 lly=0 urx=720 ury=504 rwi=7200}
\put(2485,634){\makebox(0,0)[r]{x = 0.00001}}%
\put(2485,734){\makebox(0,0)[r]{x = 0.0001}}%
\put(2485,834){\makebox(0,0)[r]{x = 0.001}}%
\put(2485,934){\makebox(0,0)[r]{x = 0.01}}%
\put(2352,476){\makebox(0,0)[l]{$d_{\perp, \pi}$}}%
\put(1590,476){\makebox(0,0)[l]{$d_{\perp, Q0}$}}%
\put(1925,50){\makebox(0,0){\Large $\dperp (fm)$}}%
\put(100,1180){%
\special{ps: gsave currentpoint currentpoint translate
270 rotate neg exch neg exch translate}%
\makebox(0,0)[b]{\shortstack{\Large $\sigma^{MFGS} (x,\dperp^2) / \sigma^{WGB}$}}%
\special{ps: currentpoint grestore moveto}%
}%
\put(3450,200){\makebox(0,0){1}}%
\put(3145,200){\makebox(0,0){0.9}}%
\put(2840,200){\makebox(0,0){0.8}}%
\put(2535,200){\makebox(0,0){0.7}}%
\put(2230,200){\makebox(0,0){0.6}}%
\put(1925,200){\makebox(0,0){0.5}}%
\put(1620,200){\makebox(0,0){0.4}}%
\put(1315,200){\makebox(0,0){0.3}}%
\put(1010,200){\makebox(0,0){0.2}}%
\put(705,200){\makebox(0,0){0.1}}%
\put(400,200){\makebox(0,0){0}}%
\put(350,2060){\makebox(0,0)[r]{2.5}}%
\put(350,1708){\makebox(0,0)[r]{2}}%
\put(350,1356){\makebox(0,0)[r]{1.5}}%
\put(350,1004){\makebox(0,0)[r]{1}}%
\put(350,652){\makebox(0,0)[r]{0.5}}%
\put(350,300){\makebox(0,0)[r]{0}}%
\end{picture}%
\endgroup
 
    \caption{The ratio of two models for ${\hat \sigma}$: the QCD model \cite{mfgs} divided by the  
saturation model \cite{wgb}.}
    \label{fig:wgbcom}
  \end{center}
\end{figure}

As an example of a comparison between models, Fig.(\ref{fig:wgbcom}) shows 
a plot of the ratio of these two models for ${\hat \sigma}$. 
The two forms have inherently different behaviour in energy and transverse 
size which is especially apparent at small and large $\dperp$ in HERA kinematics. 
For intermediate $\dperp$ the numerical values for ${\hat \sigma}$ are similar.

Forshaw, Kerley and Shaw \cite{kfs} proposed a general ansatz for 
${\hat \sigma}$,  which they stress is modelled as a function of $\dperp^2$ and $W^2$ (rather than $x$), with a soft Pomeron and a hard Pomeron piece:
\begin{equation}
{\hat \sigma} (W^2,\dperp) = a\frac{P^{2}_{s}(\dperp)}{1 + 
P^{2}_{s}(\dperp)}(\dperp^{2} W^2)^{\lambda_{s}} + \dperp^2 P^{2}_{h}(\dperp) 
e^{(-\nu_{h}^{2} \dperp)} (\dperp^{2} W^2)^{\lambda_{h}} \, , 
\label{eqkfs}
\end{equation}
\noindent where $P_{s}(\dperp)$ and $P_{h}(\dperp)$ are polynomials in $\dperp$. 
They successfully fit this general form to the inclusive data.
For very large dipoles (of the order of the size of a light meson) 
one may think that the dipole cross section should be a function of $W^2$ 
rather than $x \approx Q^2/W^2$, since it should just depend on the energy of the collision 
(here there is no hard scale with which to specify an `$x$'). 
However, in order to achieve the observed approximate Bjorken scaling
of $F_2$, it was necessary to model the dipole cross section in eq.(\ref{eqkfs}) 
as a function of the dimensionless $\dperp^2 W^2$, rather than $W^2$ alone. 
This model also gave a reasonable description of the diffractive data (see second reference of \cite{kfs}) without further tuning of parameters, but with some additional assumptions on the $q {\bar q}g$ component.

Total cross sections are related to the forward ($t=0$) part of the elastic scattering 
amplitude via the optical theorem. However, the amplitude itself is of course a 
function of $s$ and $t \equiv n_{\mu} n^{\mu} \approx -n_{\perp}^2$ (in high energy 
kinematics the momentum transfer is mainly transverse). In general, one may also consider 
the Fourier transform with respect to this momentum transfer:
\begin{equation}
A(s,t) = \int d^{2} \bperp e^{-i n_{\perp} b_{\perp}} A(s,b_{\perp}) \, .
\end{equation}
\noindent For $t = 0$ the amplitude can be represented as an integral over all 
{\it impact parameters} $b_{\perp}$, 
\begin{equation}
A(s,t=0) = \int d^{2} \bperp A(s,\bperp) \, .
\end{equation}
\noindent More generally, the cross section for a given process will also
depend on the impact parameter of the collision, a quantity that depends on both 
the scattering state and the target and should not be confused with the transverse 
diameter of the dipole, $d_{\perp}$.

Several years ago Gotsman, Levin and Maor \cite{glm} developed a form of the eikonal model 
to be applied to DIS processes
\footnote{We follow closely here the presentation in Section 2.2 of the second reference of \cite{glm}.}:
\begin{eqnarray}
{\hat \sigma} (x, \dperp) & = & \int d^{2} \bperp e^{-i n_{\perp} \dperp} \sigma (x, \dperp, \bperp) \, , \\
\sigma (x, \dperp, \bperp) & = & 2 ~( 1 - e^{-\Omega/2})  \, .
\label{eik}
\end{eqnarray}
\noindent The following convenient factorized form was assumed for the {\it eikonal} 
\begin{eqnarray}
\Omega    & = & S(\bperp) ~\frac{\pi^2}{3} ~\dperp^2 ~\alpha_s ~xg (x, \frac{4}{\dperp^2}) \, ,
\end{eqnarray}
\noindent with a Gaussian form ascribed to the normalized non-perturbative 
two-gluon form factor of the target:
\begin{eqnarray}
S(\bperp)     & = & \frac{1}{\pi R_N^2} e^{-\bperp^2/R_N^2} \, ,
\end{eqnarray}
\noindent where $R_N^2$ is related to the proton radius. It is this specific Gaussian form 
in $b_{\perp}$ which leads to the popular exponential parameterization of the $t$-dependence, 
$e^{ B_{D} t}$.
In the limit of small $\Omega$, the exponent of eq.(\ref{eik}) may be expanded and the 
familiar form of eq.(\ref{sigpqcd}) is recovered for ${\hat \sigma}$. 
As stated earlier, this corresponds to the exchange of a 
pair of gluons in the $t$-channel, treated in the leading log($Q^2$) approximation and 
constitutes the basic building block in this version of the eikonal model.
In determining whether $\Omega$ is indeed small one must examine 
the magnitude of the {\it packing factor} of partons in the cascade, defined as 
\begin{eqnarray}
\kappa = \frac{ {\hat \sigma_{pQCD}} (x, \dperp) }{\pi R_{N}^2} \, .
\end{eqnarray}
\noindent The line $\kappa = 1$ is analogous to $d_{\perp, crit} (x)$ of \cite{mfgs} and the 
critical line of \cite{wgb}. For large $\Omega$, a standard Regge analysis is adopted. 
Thus the energy dependence of this model is similar, but not identical, to the 
model described in \cite{mfgs}.

This group recently extended this ``screening corrections'' model 
to consider explicitly gluon emission in the photon wavefunction. 
For more details we urge the reader to consult the recent preprints 
\cite{gflmn,gllmt}. 

A more precise treatment of the two gluon exchange graphs relates ${\hat \sigma} (x,\dperp)$ 
to the unintegrated gluon distribution function $f(x,\lperp)$, where $\lperp$ is the 
transverse momentum in the loop below the quark box. Since 
the gluon density itself is very poorly constrained experimentally in the small-$x$ region,  the prospects for accurately constraining $f(x,\lperp)$, which is related to 
its derivative with respect to scale, from current data alone are very poor.
However, if one imposes an additional constraint on the unintegrated gluon, 
for example that it obeys the CCFM equation \cite{ccfm} then the distribution 
becomes more constrained (see \cite{kimber} for a recent attempt to extract 
$f(x,k_{\perp})$ along these lines). Unfortunately, this involves making a 
set of choices on precisely how to do this and hence introduces a 
certain additional model dependence into the answer.

To summarize, each of the four models for ${\hat \sigma}$ that we discussed contain a perturbative 
piece that rises steeply with energy at small $\dperp$ and a part which rises much more slowly with 
energy that dominates at large transverse distances. In HERA kinematics, in each of the models the 
short distance part of ${\hat \sigma}$ reaches magnitudes typical of those observed in hadronic 
interactions (tens of millibarns). The rapid rise of this piece with energy is tamed in three of the models we discussed \cite{mfgs,glm,wgb}. 

\section*{Acknowledgments}

It is a pleasure to acknowledge the financial support of the DESY directorate.
The Amirim meeting was partially supported by the U.S.-Israel Binational 
Science Foundation (BSF) and the Israel Science Foundation (ISF). I would like 
to thank the organizors for inviting me and for generous assistance with 
travel money. I am grateful to the participants of the meeting for stimulating and lively discussions and for making important contributions to this summary.

%List of participants and affiliations: Tel Aviv University, H.~Abramowicz,  J.~Bartels, L.~Frankfurt, K.~Golec-Biernat, E.~Gurvich, H.~Jung, S.~Kananov, A.~Kreisel, H.~Kowalski, E.~Levin, A.~Levy, P.~Marage,  U.~Maor,  E.~Naftali, P.~Saull, S.~Schlenstedt, G.~Shaw, M.~Strikman, K.~Tuchin and J.~Whitmore. 

List of participants and affiliations: H.~Abramowicz, L.~Frankfurt, 
E.~Gurvich, S.~Kananov, A.~Kreisel, E.~Levin, A.~Levy, U.~Maor, E.~Naftali, 
K.~Tuchin (Tel Aviv University), P.~Saull (Tel Aviv University and Penn State University), J.~Bartels, K.~Golec-Biernat (Universit\"{a}t Hamburg), H.~Jung (Lund University), H.~Kowalski (DESY, Hamburg), S.~Schlenstedt (DESY, Zeuthen),  
P.~Marage (Universit\'{e} Libre de Bruxelles), M.~McDermott, G.~Shaw 
(Manchester University), M.~Strikman, J.~Whitmore (Penn State University).

\end{document}